# Very high upper critical fields in MgB$_2$ produced by selective tuning of impurity scattering.

A. Gurevich[§], S. Patnaik[*], V. Braccini[§,***], K. H. Kim[**], C. Mielke[**], X. Song[§], L. D. Cooley[#] S. D. Bu[§], D. M. Kim[§], J. H. Choi[§], L. J. Belenky[§], J. Giencke[§], M. K. Lee[§], W. Tian[&], X.Q. Pan[&], A. Siri[***], E. E. Hellstrom[§], C. B. Eom[§], D.C. Larbalestier[§].

[§] University of Wisconsin, Madison WI 53706, USA
[*]Now at School of Physical Sciences, Jawaharlal Nehru University, New Delhi 110067 India
[**]National High Magnetic Field Laboratory, Los Alamos National Laboratory, Los Alamos NM 87545, USA
[***] I.N.F.M.-LAMIA, Department of Physics, Via Dodecaneso 33, 16146 Genova, Italy
[#] Now at Brookhaven National Laboratory, Upton NY, USA.
[&] Department of Materials Science and Engineering, University of Michigan, Ann Arbor, MI 48109, USA

We report a significant enhancement of the upper critical field H$_{c2}$ of different MgB$_2$ samples alloyed with nonmagnetic impurities. By studying films and bulk polycrystals with different resistivities $\rho$, we show a clear trend of H$_{c2}$ increase as $\rho$ increases. One particular high resistivity film had zero-temperature $H_{c2}(0)$ well above the $H_{c2}$ values of competing non-cuprate superconductors such as Nb$_3$Sn and Nb-Ti. Our high-field transport measurements give record values $H_{c2}^{\perp}(0) \approx 34$T and $H_{c2}^{\parallel}(0) \approx 49$ T for high resistivity films and H$_{c2}$(0) $\approx$ 29 T for untextured bulk polycrystals. The highest H$_{c2}$ film also exhibits a significant upward curvature of H$_{c2}$(T), and temperature dependence of the anisotropy parameter $\gamma(T) = H_{c2}^{\parallel}/ H_{c2}^{\perp}$ opposite to that of single crystals: $\gamma(T)$ *decreases* as the temperature decreases, from $\gamma(T_c) \approx 2$ to $\gamma(0) \approx 1.5$. This remarkable H$_{c2}$ enhancement and its anomalous temperature dependence are a consequence of the two-gap superconductivity in MgB$_2$, which offers special opportunities for further $H_{c2}$ increase by tuning of the impurity scattering by selective alloying on Mg and B sites. Our experimental results can be explained by a theory of two-gap superconductivity in the dirty limit. The very high values of H$_{c2}$(T) observed suggest that MgB$_2$ can be made into a versatile, competitive high-field superconductor.

## 1. Introduction

Since the discovery of the "intermediate-$T_c$" superconductor MgB$_2$ with critical temperature $T_c$ of 39 K [1,2], many efforts have been made to increase the ability of MgB$_2$ to sustain superconductivity at higher magnetic fields. This problem is of fundamental interest and is vital for applications because MgB$_2$ exhibits high critical current densities $J_c$ [3-6], no intrinsic current blockage by grain boundaries [7], and comparatively weak anisotropy and thermal fluctuations. Yet, clean MgB$_2$ has so far offered no clear advantages as compared to existing practical high-field superconductors because of rather low upper critical fields, $H_{c2}^{\perp}(0) \approx 3.5$ T and $H_{c2}^{\parallel}(0) \approx 18$ T perpendicular and parallel to the *ab* plane of MgB$_2$ single crystals [8-12]. Indeed, the key to magnet applications of superconductors lies in the rare combination of low cost and a ready wire fabrication route, which produces conductors with high upper critical fields $H_{c2}$, and critical current densities $J_c$ [13]. Modern conductors are made from Nb-Ti ($T_c = 9$ K, $H_{c2}(4.2$ K$) = 10$ T), Nb$_3$Sn ($T_c = 18$ K, $H_{c2}(4.2$ K$) = 28$ T), and the high-temperature superconductors (HTS), Bi$_2$Sr$_2$CaCu$_2$O$_{8-x}$ ($T_c \sim 90$ K), and (Bi,Pb)$_2$Sr$_2$Ca$_2$Cu$_3$O$_{10-x}$ ($T_c \sim 110$ K). At low temperatures the HTS polycrystals have $H_{c2}$ well above 50 T, but they suffer from strong current relaxation, high anisotropy ($H_{c2}^{\parallel}/H_{c2}^{\perp} > 20$), and weak-linked grain boundaries that obstruct the current except when crystallographically textured. Because of their large $H_{c2}$ anisotropy, HTS conductors must be made in tape form, whereas isotropic Nb-based superconductors are easily fabricated into round wires capable of generating magnetic fields above 20 T, but only at $T = 2$-$5$ K. However, due to recent advances in cryocoolers, many electric utility, fusion and high-energy physics applications may be best optimized at temperatures of 10 to 35 K, [14] a domain for which MgB$_2$ could provide the cheapest superconducting wires [15]. It is the low upper critical field which has mainly limited the potential applications of clean MgB$_2$, for which $H_{c2}(4.2$ K$)$ is even less than that of Nb-Ti. Here we show how $H_{c2}(T)$ of MgB$_2$ can be radically enhanced beyond that of any Nb-based superconductor by introducing strong impurity scattering. The resulting increase of $H_{c2}$ along with the significant decrease of H$_{c2}$ anisotropy, from $H_{c2}^{\parallel}/H_{c2}^{\perp} \approx 5$-$6$ [2] down to $H_{c2}^{\parallel}/H_{c2}^{\perp} < 2$, can make MgB$_2$ well suited for high-field magnet applications at 20-30 K where HTS conductors presently are without challenge [13].

We show that a significant portion of the anomalous increase in $H_{c2}$ results from the two-gap superconductivity in MgB$_2$, which has been well established by many *ab initio* calculations [16,17], STM experiments [18], point contact [19] and Raman [20] spectroscopy, specific heat [21] and neutron diffraction [22] measurements. As a result, there is compelling evidence that MgB$_2$ is the first superconductor with two weakly coupled superconducting gaps $\Delta_\sigma(4.2$ K$) \approx 7.2$ meV and $\Delta_\pi(4.2$ K$) \approx 2.3$ meV. These gaps reside on different disconnected sheets of the Fermi surface, which is comprised of nearly cylindrical parts formed by *in-plane* $\sigma$ antibonding p$_{xy}$ orbitals of B, and a more isotropic tubular network formed by *out-of-plane* $\pi$ bonding and antibonding p$_z$ orbitals of B. Such a Fermi surface gives rise to three different impurity scattering channels in MgB$_2$ alloys: intraband scattering within each $\sigma$ and $\pi$ sheet and interband scattering between them [23]. It is these multiple scattering channels which make it possible to increase $H_{c2}$ of MgB$_2$ to a much greater extent than in one-gap superconductors, not only by the usual increase of the normal state resistivity $\rho$, as in low-$T_c$ superconductors [24-26], but also by optimizing the relative weight of the $\sigma$ and $\pi$ scattering rates by selective substitution of boron or magnesium.

In low-$T_c$ s-wave superconductors, alloying with nonmagnetic impurities does not affect $T_c$ much but does increase the slope $H_{c2}' = |dH_{c2}/dT|$ at $T_c$ proportionally to the residual resistivity $\rho$. This behavior is characteristic of the dirty limit $\hbar v_F \gg 2\pi l k_B T_c$, where $l$ is the electron mean free path, $v_F$ is the Fermi velocity, ℏ and $k_B$ are the Planck and Boltzmann constants, respectively. The zero-temperature field $H_{c2}(0)$ is then given by [26]

$$H_{c2}(0) = 0.69 T_c H'_{c2}(T_c) \qquad (1)$$

Because $H_{c2}(0)$ increases proportionally to ρ, alloying is a well-established route to $H_{c2}$ enhancement in Nb-Ti [24] and the A15 compounds [25]. The same approach has been successfully applied to $MgB_2$ as well, which was demonstrated by alloying of $MgB_2$ films [27], bulk samples [28], and also by proton [29] and neutron [30] irradiation of polycrystalline $MgB_2$. In what follows we show, both experimentally and theoretically, that because of the two-gap superconductivity in $MgB_2$, Equation (1) in fact strongly *underestimates* the actual $H_{c2}(0)$ of dirty $MgB_2$.

## 2. Experimental

To show how far $H_{c2}$ can be actually increased by alloying, we performed high-field transport measurements on samples of very different purity: a fiber-textured, high-resistivity (ρ(40K) = 220 μΩ-cm) c-axis-oriented film on (111) $SrTiO_3$ substrate [31], a low-resistivity (ρ(40K) = 7 μΩ-cm), c-axis-oriented, epitaxial $MgB_2$ film on (0001) $Al_2O_3$ substrate [32], and 3 untextured polycrystalline bulk $MgB_2$ samples of different resistivities, ρ(40K) = 1-18 μΩ-cm [33]. The 220 μΩcm, pulsed laser deposited film was earlier reported to have a macroscopically inhomogeneous resistivity ρ(40K) ≈ 360 - 440 μΩcm [27,31], but subsequent transmission electron microscopy [34] showed that approximately a half of the measured film thickness was a fully reacted $MgB_2$. The film had $T_c \sim 31K$, and extremely high resistivity compared to the nominal ρ(40K) ≈ 1μΩcm of "clean" $MgB_2$ [12]. The sputtered epitaxial film [32] has 400 nm of $MgB_2$ thickness, $T_c = 35$ K and much lower ρ(40K) ≈ 7μΩcm. The bulk samples have ρ(40K) = 1 μΩcm as made by direct reaction, 18 μΩcm after subsequently heating in Mg vapor followed by slow cooling [33]. After aging for two months, the resistivity ρ(40K) for this sample fell from 18 μΩcm down to 5 μΩcm.

Measurements of $H_{c2}(T)$ on thin films were made partly in DC fields up to 14 T in a Quantum Design Physical Property Measurement System (PPMS) and partly in the high-field facility at LANL capable of generating 32 ms asymmetric field pulses up to 50 T. A combination of an ultra fast digitizer and a lock-in amplifier enabled us to take over 32,000 data points of the measured 100 kHz signal during the pulse, from $T_c$ to 1.4 K with ~ 10 mK sensitivity. Three data sets taken on the films gave consistent results. The greater cross-section ~ 0.5 × 0.5 mm and lower ρ of the bulk samples made the resistive transitions under pulsed fields so noisy that we had to measure them under dc conditions. Dc measurements up to 14 T were performed in the PPMS while high field resistance $R(B)$ measurements were made in the 33T resistive magnet at the NHMFL in Tallahassee Florida on the sample annealed in Mg vapor after it was aged for two months.

Figures 1a and 1b show the resistive transitions from 1.5 to 30K of the 220 μΩcm film as a function of pulsed magnetic fields for perpendicular and parallel fields. These transitions agree well with our earlier dc measurements up to 9T of another piece of this same film for which we defined effective irreversibility field $H^*(T)$ as $R(H^*) = 0.1R(T_c)$, and the upper critical field as $R(H_{c2}) = 0.9R(T_c)$ [27]. We use here a more conservative criterion for $H_{c2}$ in Figures 1 than the usual definition for $H_{c2}$ as the onset of the resistive transitions, which are much narrower ($H^*(T) \sim 0.8$-$0.9H_{c2}(T)$) as compared to those observed on high-$T_c$ cuprates [35]. Although our definition underestimates the actual upper critical field (as seen from Figure 1, the field differences between the onset and the $0.9R(T_c)$ point can be as high as 3-4 Tesla), it makes it less sensitive to the noise on the upper parts of the $R(H)$ curves.

$H_{c2}^{\parallel}(T)$ and $H_{c2}^{\perp}(T)$ curves for the 2 films are shown in Figure 2, from which it is immediately clear that the temperature dependence of $H_{c2}(T)$ is rather anomalous. Indeed, Equation (1) would predict $H_{c2}^{\perp}(0) \approx 10.5$ T for the cleaner film, while in fact it exceeds 20 T. In the parallel configuration, Equation (1) gives $H_{c2}^{\parallel}(0) \approx 20$ T versus the measured 30 T. Even more striking is the behavior of the dirtier film. Record high values of $H_{c2}^{\perp}(0) \approx 34$ T and $H_{c2}^{\parallel}(0) \approx 49$ T are attained, both considerably exceeding the earlier low-field extrapolations of 20 and 39 T based on measurements close to $T_c$ and the use of Equation (1) [27]. Both parallel and perpendicular $H_{c2}(T)$ curves exhibit *upward* curvature and different shapes of $H_{c2}(T)$ curves for parallel and perpendicular field orientations, neither being consistent with the one-gap theory [26]. We also show the field at which $R(H) = 0.1R(T_c)$ to emphasize that the extreme high field performance capability of dirtier $MgB_2$ is not very sensitive to the particular definition of $H_{c2}$. Moreover, for $H \parallel ab$, $H_{c2}$ of the dirty $MgB_2$ significantly exceeds $H_{c2}$ of the best $Nb_3Sn$ at all temperatures. Even for the perpendicular field ($H \perp ab$), not only is $H_{c2}(T)$ of $MgB_2$ still superior to $Nb_3Sn$, but dirty $MgB_2$ has essentially the same $H_{c2} = 12$ T at 20 K as the best Nb-Ti at 4.2 K.

Figure 3 collects the data for bulk, untextured $MgB_2$ and again shows the decisive influence of resistivity. The low-resistivity ($\rho(40 \text{ K}) \approx 1$ μΩ-cm) sample has an extrapolated $H_{c2}(0) \sim 17$ T, close to that of other clean limit samples [12]. After exposure to Mg, the resistivity $\rho(40K)$ rose to 18 μΩ-cm and $H_{c2}' = 1.2$ T/K at 25-30 K became almost as high as $H_{c2}'$ for 220 μΩ-cm film in a perpendicular field. $H_{c2}$ then reached 9 T at 27 K, comparable to the film value in parallel field. After remeasuring this sample about two months later, it had apparently aged, as evidenced by a reduced $\rho = 5$ μΩ-cm, a decreased $H_{c2}(T)$, and an increased $T_c$ from 36.9 to 37.7 K. This aging may result from relaxation of stress in the quenched crystalline structure. In any case, the $H_{c2}(T)$ data show the same qualitative trends as the film data in Figure 2: as the resistivity increases from 1 to 5 μΩ-cm, $H_{c2}(0)$ approximately doubles, reaching $\sim 29$ T, close to the best $H_{c2}(0)$ value of $Nb_3Sn$. We do note that the interpretation of resistive transitions for polycrystals may be complicated by the percolative effects due to the anisotropy of $H_{c2}$, which results in different resistive transitions for crystallites with varying orientations with respect to the applied field.

The totality of the data presented in Figures 2 and 3 unambiguously indicates that impurity scattering can markedly enhance $H_{c2}$ of $MgB_2$ while only weakly reducing $T_c$. For example, as $\rho(40 \text{ K})$ is increased from $\sim 1$ μΩ-cm to $\sim 220$ μΩ-cm, the perpendicular field $H_{c2}^{\perp}(0)$ rises nearly 10 fold, from the typical single crystal values of $\approx 3$-5 T to $\sim 35$-39 T. However this $H_{c2}$ enhancement is not entirely determined by the resistivity, as in dirty one-gap superconductors. As follows from Figure 2, a significant portion of the $H_{c2}$ increase comes from the anomalous low temperature upward curvature of $H_{c2}(T)$.

### 3. Discussion

#### a) Two-gap, dirty limit theory

To address the physics behind the observed anomalous behavior of $H_{c2}(T)$, we applied a dirty-limit two-gap superconductivity theory based on the Usadel equations [36,37] to $MgB_2$. In this approach the details of the complex Fermi surface of $MgB_2$ are not essential for the calculation of $H_{c2}$, while the impurity scattering is accounted for by the normal state electronic diffusivity tensors $D_{\sigma}^{\alpha\beta}$ and $D_{\pi}^{\alpha\beta}$, controlled by their respective intraband scattering rates in the σ and π bands, and , and interband scattering rates $\Gamma_{\pi\sigma}$ and $\Gamma_{\sigma\pi}$. For MgB$_2$, the interband scattering (responsible for $T_c$ suppression by nonmagnetic impurities) is weak [23], so we assume for simplicity that $\Gamma_{\pi\sigma} = \Gamma_{\sigma\pi} = 0$ (a more general case of finite $\Gamma_{\pi\sigma}$ and $\Gamma_{\sigma\pi}$ was considered in Ref. 36). In that case the solution of the linearized two-gap Usadel equations for H⊥ab and $\Gamma_{\pi\sigma} = \Gamma_{\sigma\pi} = 0$, gives the following equation for $H_{c2}^\perp$ [36,37]

$$2w[\ln t + u(b/t)][\ln t + u(\eta b/t)] + \lambda_2[\ln t + u(\eta b/t)] + \lambda_1[\ln t + u(b/t)] = 0. \quad (2)$$

Here $t = T/T_c$, $u(b) = \psi(1/2 + b) - \psi(1/2)$, $\psi(x)$ is the Euler digamma function, $b = \hbar H_{c2} D_\sigma / 2\phi_0 k_B T_c$, $\phi_0$ is the magnetic flux quantum, $\eta = D_\pi/D_\sigma$, $w = \lambda_{\sigma\sigma}\lambda_{\pi\pi} - \lambda_{\sigma\pi}\lambda_{\pi\sigma}$, $\lambda_{1,2} = \lambda_0 \pm \lambda_-$, $\lambda_0 = (\lambda_-^2 + 4\lambda_{\sigma\pi}\lambda_{\pi\sigma})^{1/2}$, $\lambda_- = \lambda_{\sigma\sigma} - \lambda_{\pi\pi}$, and the 2 × 2 matrix of the BCS superconducting coupling constants $\lambda_{mn} = \lambda_{mn}^{(ep)} - \mu_{mn}$ contains both the electron-phonon constants $\lambda_{mn}^{(ep)}$ and the Coulomb pseudopotential $\mu_{mn}$. First principles calculations of the intraband ($m = n$) and interband ($m \neq n$) matrix elements $\lambda_{mn}$ for MgB$_2$ gave $\lambda_{\sigma\sigma} = 0.81$, $\lambda_{\pi\pi} = 0.28$, $\lambda_{\sigma\pi} = 0.115$ and $\lambda_{\pi\sigma} = 0.09$ [38].

The $H_{c2}(T)$ curves calculated from Equation (2) evolve from the classic dirty-limit, one-band BCS behavior [26] at $D_\pi = D_\sigma$ to rather different $H_{c2}(T)$ curves which have portions with both upward and downward curvatures for either $D_\pi \ll D_\sigma$ and $D_\pi \gg D_\sigma$ [36]. As a result, $H_{c2}(0)$ can be significantly higher than suggested by the conventional extrapolation (1). To show how this happens, we first obtain $H_{c2}(T)$ near $T_c$ by expanding Equation (2) using $u(b) \approx \pi^2 b/2$ at $b \ll 1$:

$$H_{c2}^\perp(T) = \frac{8\phi_0 \lambda_0 k_B (T_c - T)}{\hbar \pi^2 (\lambda_1 D_\sigma + \lambda_2 D_\pi)} \quad (3)$$

For the characteristic $\lambda_{mn}$ values of MgB$_2$, but very different $D_\pi$ and $D_\sigma$, the upper critical field is therefore determined by the maximum intraband diffusivity, for example, by $D_\pi$ if disorder mostly causes scattering in the "strong" σ band. The zero-temperature value $H_{c2}(0)$ is obtained using the limiting behavior of $u(b/t) = \ln(4\gamma b/t)$ for $t \ll 1$ where $\ln\gamma = -0.577$. Then Equation (3) reduces to a quadratic equation for $\ln H_{c2}$, whence

$$H_{c2} = \frac{\phi_0 k_B T_c}{2\gamma \hbar \sqrt{D_\pi D_\sigma}} \exp(g), \quad (4)$$

where $2g = [(\lambda_0/w)^2 + \ln^2\eta + 2\ln(\eta)\lambda_-/w]^{1/2} - \lambda_0/w$. For very different diffusivities, Equation (4) yields $H_{c2}(0) \approx \phi_0 k_B T_c \exp(-\lambda_2/2w)/2\gamma\hbar D_\sigma$ if $D_\sigma \ll D_\pi$, but $H_{c2}(0) \approx \phi_0 k_B T_c \exp(-\lambda_1/2w)/2\gamma\hbar D_\pi$ if $D_\pi \ll D_\sigma$, so $H_{c2}(0)$ diverges as either $D_\sigma$ or $D_\pi$ goes to zero. Thus, unlike the region $T \approx T_c$ where $H_{c2}(T)$ is determined by the *maximum* diffusivity between $D_\sigma$ and $D_\pi$, the zero-temperature $H_{c2}(0)$ is controlled by the *minimum* diffusivity. It is this feature of two-gap superconductivity, which causes both the upward curvature of $H_{c2}(T)$ and the violation of Equation (1) for $D_\pi \neq D_\sigma$. Equation (2) can be generalized to arbitrary field orientation by replacing both intraband diffusivities $D_m$ with the angular dependent effective diffusivities [36]

$$D_m(\theta) = (D_m^{(ab)2}\cos^2\theta + D_m^{(c)}D_m^{(ab)}\sin^2\theta)^{1/2}, \quad (5)$$

where θ is the angle between **H** and the c-axis, $m = (\sigma, \pi)$, and the superscripts (ab) and (c) indicate the diffusivities along the ab plane and the c-axis, respectively. Equations (2) and (3) enable us to address the observed anomalous temperature dependence of the anisotropy of $H_{c2}(\theta)$ [8-12], as described in detail elsewhere [36]. Here we only use the fact that for the parallel field orientation, $H\|ab$, the effective diffusivities in Equations (2)-(4) become $D_m = [D_m^{(ab)}D_m^{(c)}]^{1/2}$.

Equation (2) was used to describe the observations of Figure 2. We focus here on the most resistive film for which the theory describes well all anomalous features of the observed dependence of $H_{c2}(T)$, including the upward curvature and the marked increase of $H_{c2}(T)$ at low temperatures. The good agreement between the theory and experiment not only gives one more argument for the two-gap scenario in $MgB_2$, but also suggests a new way to increase $H_{c2}$ by manipulation of the diffusivity ratio $D_\sigma/D_\pi$ using selective atomic substitutions on either Mg or B sites. Both the ratio $D_\sigma/D_\pi$ and the anisotropy parameter $D_m^{(ab)}/D_m^{(c)}$ for this dirty film can be extracted from the fit shown in Figure 2, giving $D_\pi^{(ab)} \approx 0.12 D_\sigma^{(ab)}$ for $H\perp ab$ and $[D_\pi^{(ab)}D_\pi^{(c)}]^{1/2} \approx 0.2[D_\sigma^{(ab)}D_\sigma^{(c)}]^{1/2}$ for $H\|ab$. This relation $D_\pi^{(ab)} \sim 0.1 D_\sigma^{(ab)}$ indicates that the π-scattering in the dirty film is much stronger than the σ-scattering. Thus, according to Equations (3) and (4), the slope of $H_{c2}(T)$ near $T_c$ is determined by the σ-scattering most likely is due to impurities or disorder on B sites, whereas the additional increase of $H_{c2}$ at $T \ll T_c$ results from the predominantly π-scattering on Mg sites.

### b) $H_{c2}$ anisotropy

As follows from Figure 2, the shape of $H_{c2}(T)$ curves change as $H$ rotates from $H\perp ab$ to $H\|ab$. This behavior, inconsistent with the one-gap Ginzburg-Landau theory, is also described well by Equations (2) and (5), in which the orientational dependence of $H_{c2}$ is mostly controlled by the highly anisotropic diffusivity $D_\sigma(\theta)$ for the nearly 2D σ band. For the dirty film, the fit described above gives $D_\pi^{(ab)} \approx 0.12 D_\sigma^{(ab)}$ for $H\perp ab$ and $[D_\pi^{(ab)}D_\pi^{(c)}]^{1/2} \approx 0.2[D_\sigma^{(ab)}D_\sigma^{(c)}]^{1/2}$ for $H\|ab$, whence we obtain $D_\sigma^{(ab)}/D_\sigma^{(c)} \approx 3 D_\pi^{(ab)}/D_\pi^{(c)}$. Thus, the diffusivity for the nearly 2D σ-band is more anisotropic than the diffusivity for the 3D π-band, in qualitative agreement with *ab-initio* Fermi surface calculations [16,17] and STM measurements [18], which do show weaker tunneling probability to the σ-band along the c-axis. However, our estimate of the ratio $D_\sigma^{(ab)}/D_\sigma^{(c)} \approx 3$ for the dirty film is smaller than the ratio of the Fermi velocities, $v_F^{(ab)} \sim 10^2 v_F^{(c)}$ for the σ band inferred from *ab-initio* calculations for clean $MgB_2$ single crystals [16,17]. This discrepancy might be due to the significant buckling of the boron planes observed by TEM on our dirty films (see below). Such buckling increases both the out-of-plane electron scattering and the σπ band hybridization, reducing the initial anisotropy of the σ band in the dirty film.

The temperature dependence of the anisotropy parameter $H_{c2}^{\|}/H_{c2}^{\perp}$ extracted from the data of Figure 2 for both the high-resistivity and epitaxial films is shown in Figure 4. As $T$ decreases, the ratio $H_{c2}^{\|}/H_{c2}^{\perp}$ in both cases *decreases*, from $\approx 2$ at $T_c$ down to $\approx 1.5$ at 0 K. This behavior is opposite to the usual temperature dependence of $H_{c2}(T)$ in $MgB_2$ single crystals for which the anisotropy parameter $\gamma(T) = H_{c2}^{\|}/H_{c2}^{\perp}$ typically *increases* from $\approx 3$ at $T_c$ up to 5-6 at 0 K [8-12]. This unusual dependence of $\gamma(T)$ for the dirty film is due to the large difference between σ and π electron diffusivities ($D_\pi \sim 0.1 D_\sigma$). As a result, the anisotropy parameter $\gamma(T)$ near $T_c$ is determined by cleaner, more anisotropic 2D σ band. However, at low temperatures the anisotropy of $H_{c2}$ decreases because $\gamma(T)$ is mostly determined by the dirtier 3D π band. Thus, the theory based on Equations (2)-(5) does explain the observed decrease of the anisotropy ratio $H_{c2}^{\|}/H_{c2}^{\perp}$ as $T$ decreases. The anomalous temperature dependence of $H_{c2}^{\|}/H_{c2}^{\perp}$ is one more manifestation of the two-gap superconductivity in $MgB_2$. Interestingly, the much less resistive epitaxial film also exhibits a qualitatively similar behavior of $\gamma(T)$.

### c) Impurity scattering mechanisms

At present the mechanisms of impurity scattering in alloyed $MgB_2$ and its complex substitutional chemistry [12,39] are not well understood, though we do know that the previously studied [31] quarter of our dirty 220 µΩ-cm film was oxygen-rich and had a Mg:B:O ratio of 1:0.9:0.7 as measured by wavelength dispersive spectroscopy [34]. Both O and C, which substitute for B, can result in strong σ-scattering [23]. A further insight can be inferred from the electron micrographs presented in Figures 5a and 5b, which show the "brick-wall" grain nanostructure of this film, which stands in stark contrast to the structure of the lower resistivity epitaxial film [32], which has almost perfect layered atomic structure, as shown in Figure 5c. By contrast, the dirty film has a 10-20 nm $MgB_2$ grain size and about 20% of MgO particles of similar size, which produce much distortion of the atomic packing and thus causes significant scattering in both σ and π bands. Another important feature of the dirty film shown in Figure 5b is a pronounced buckling of the Mg planes along the c-axis, which manifests itself both in variations of the Mg-B spacing and in the intensity of Mg spots. The Mg-plane buckling certainly gives rise to strong out-of-plane π scattering, which may explain the relation $D_\pi^{(ab)} \sim 0.1 D_\sigma^{(ab)}$ obtained from the fit in Figure 2.

In the above analysis of the $H_{c2}$ data we assumed that elastic scattering by point defects (nonmagnetic impurities) is the dominant mechanism of the residual resistivity ρ. In this case ρ can be expressed in terms of the diffusivities $D_\sigma$ and $D_\pi$ by

$$1/\rho = e^2(N_\sigma D_\sigma + N_\pi D_\pi), \qquad (6)$$

where $N_\sigma$ and $N_\pi$ are partial densities of states in the σ and π bands, respectively, $e$ is the electron charge, and the weak interband scattering is neglected. Using Equations (3) and (6), $D_\sigma$ and $D_\pi$ can be extracted from the observed ρ and $H_{c2}'$ at $T_c$ [36]. However, in addition to impurity scattering, the resistivity of real $MgB_2$ samples can also be strongly affected by extrinsic mechanisms, such as scattering on grain boundaries or second phase nanoprecipitates, which effectively reduce the current-carrying cross-section [40]. While it is not easy to unambiguously separate extrinsic and intrinsic contributions to ρ, we can nevertheless make qualitative conclusions about the scattering mechanisms by analyzing *both* $H_{c2}$ and ρ data. Indeed, $H_{c2}$ can be only affected by disorder on the scale shorter than the clean limit GL coherence length, $\xi_\sigma(0) = [\phi_0/2\pi H_{c2}(0)]^{1/2}$, which is about 8-10 nm for the values of $H_{c2}(0) \approx 3.5 - 5$ T characteristic of single crystals [8-12]. Therefore, the fact that $H_{c2}$ of the dirty film is increased by almost 10 times as compared to single crystals does indicate strong scattering on the atomic scale << 10 nm, most likely due to impurities, especially O, on the B and Mg planes and quenched crystalline disorder and buckling of Mg planes, as discussed above. In turn, the grain structure on the scale 10-20 nm > $\xi_\sigma(0)$ clearly seen in Figure 5a may not be the primary reason for the observed $H_{c2}$ enhancement, although additional scattering on grain boundaries and the MgO precipitates can certainly increase the global resistivity.

Our data clearly show the decisive qualitative effect of resistivity on $H_{c2}$, however the extrinsic mechanisms [40] may be indeed very important for understanding why $H_{c2}$ even in dirty $MgB_2$ samples does not always scale with ρ and why samples with comparable $H_{c2}$ can have very different ρ. For instance, our dirty 220 µΩcm film has nearly the same slope $H_{c2}'$ at $T_c$ as the 18 µΩ-cm Mg-annealed polycrystal. We can therefore conclude that scattering on the nanoscale grain boundary structure shown in Figure 5 could indeed cause a significant extrinsic contribution $\rho_{ex}$ to the overall resistivity ρ. To account for $\rho_{ex}$ in the above dirty limit theory, one should substitute only the impurity part, ρ - $\rho_{ex}$, instead of ρ in Equation (6) which determines the electron diffusivities $D_\sigma$ and $D_\pi$. Since the extrinsic mechanisms do not affect $H_{c2}$, and they do not change results of our analysis based on Equations (2)-(6).

In principle, the contribution of $\rho_{ex}$ to the total resistivity $\rho$ could be extracted from the measured $\rho(T_c)$ and $H_{c2}(T)$ curve with the use of Equations (2) and (6). First, from the fit to $H_{c2}(T)$ curve, one can evaluate the values of $D_\sigma$ and $D_\pi$, using the coupling constants $\lambda_{mn}$ and the partial densities of states $N_\sigma$ and $N_\pi$ from *ab-initio* calculations [16,17,38] as input parameters in Equation (2). Next, the impurity contribution $\rho_i$ can be calculated, upon substitution of the so-obtained $D_\sigma$ and $D_\pi$ into Equation (6), giving the extrinsic contribution, $\rho_{ex} = \rho - \rho_i$.

### d) MgB$_2$ as an emerging high-field conductor material

The observed significant enhancement of $H_{c2}(T)$ both for bulk samples and especially for the high-resistivity thin film may provide the missing component to the other essentials needed for a viable MgB$_2$ conductor technology. Low raw material cost, round wire capability, high critical current densities and irreversibility fields are all possessed by MgB$_2$ today [6]. Our results show that the two-band physics of dirty MgB$_2$ provides a new way to boost $H_{c2}$ at low temperatures, so that it betters all low-$T_c$ superconductors except PbMo$_6$S$_8$, for which $H_{c2}(0) = 59$ T, as shown in Figure 6. Since no viable wire route has ever been developed for PbMo$_6$S$_8$, the real competitors to MgB$_2$ are only HTS tapes of Bi$_2$Sr$_2$CaCu$_2$O$_{8-x}$, (Bi,Pb)$_2$Sr$_2$Ca$_2$Cu$_3$O$_{10-x}$, YBa$_2$Cu$_3$O$_{7-\delta}$ or Nb$_3$Sn or Nb-Ti [3]. As follows from Figure 6, MgB$_2$ is already competitive with Nb$_3$Sn at low temperatures and with cuprate HTS at 20-25 K for electric utility applications for which fields up to ~ 5 T are needed [13]. Today the majority of superconducting magnet applications use Nb-Ti at 4.2 K and fields below 7 Tesla. This field capability is now open to MgB$_2$ at much higher temperatures 15-25 K accessible to cryocoolers where the wide stability margin of MgB$_2$ can offer clear advantage over all Nb-based superconductors [14].

One important path for making suitable MgB$_2$ conductors is based on the powder-in-tube (PIT) route to both round wires and tapes [6]. This process has been validated by the production of high-performance, round PIT Nb$_3$Sn wires, which typically have several hundreds of 20-50 μm filaments with low hysteretic loss [41]. Although tape-like HTS conductors are commercially available, they have relatively high ac losses and cannot easily be cabled to make conductors of arbitrary amperage. By contrast, MgB$_2$ offers the advantages of round-wire geometry, low raw materials cost, high strength, capability of being fabricated by a proven wire technology coupled with high $J_c$ and moderate $H_{c2}$ anisotropy. As we show in this paper, strong π scattering due to disorder in the Mg sublattice can significantly boost the upper critical field at low temperatures without great penalty to $T_c$. Another important result is that introducing strong scattering in the π band by alloying the Mg sublattice can significantly reduce the $H_{c2}$ anisotropy, as compared to MgB$_2$ single crystals (see Figure 4). We believe that the peculiar high-field behavior of our samples and their good description by a theory of dirty two-gap superconductivity clearly show the general trend for further $H_{c2}$ enhancement in MgB$_2$. Therefore, further systematic studies of the unexplored physics and materials science of MgB$_2$ alloys combined with high-field $H_{c2}$ and $J_c$ measurements would be invaluable both for better understanding of high-field superconductivity in MgB$_2$ and for future magnet applications.

The results of this work also pose a fundamental question: How far can alloying further increase $H_{c2}$ of MgB$_2$? As we have shown both experimentally and theoretically, the two-band physics does remove one essential obstacle for higher $H_{c2}$ at low $T$ due to replacing the factor 0.69 in Equation (1) by a function $\chi(D_\sigma, D_\pi)$, which can be *greater than 1*. As a result, much higher $H_{c2}(0)$ values are possible for a given slope $H_{c2}'$ at $T_c$. To properly understand what further improvements in properties can be derived by alloying MgB$_2$, it is vital to measure the whole $H_{c2}(T)$ curve over a wide range of T and ρ, so that the peculiar alloying potential of this apparently simple material with complex substitutional chemistry [39] can be understood. For instance, for dirty MgB$_2$ with values of $H_{c2}'$ = 1 T/K, $T_c$ = 40 K, and $\chi$ = 1, the theory predicts $H_{c2}(0)$ = 40 Tesla. Such a high $H_{c2}(0)$ already greatly exceeds $H_{c2}(0)$ of Nb$_3$Sn, even though $H_{c2}'$ is still significantly smaller than the 2 T/K characteristic of other LTS and of HTS, not to mention ~ 2.5 T/K for PbMo$_6$S$_8$ (see Figure 6). For $H_{c2}'$ = 1 T/K, the shortest GL in-plane coherence length $\xi_\sigma(0) = [\phi_0/2\pi T_c H_{c2}']^{1/2} \approx 3$ nm for the σ band is still large enough to ensure no significant magnetic granularity and weak link behavior at grain boundaries. Thus, there are no inherent limitations to further increase of $H_{c2}'$ toward the HTS, Nb-Ti and Nb$_3$Sn levels of 2 T/K by proper alloying with nonmagnetic impurities or by quenched-in lattice disorder in MgB$_2$. For $H_{c2}' \approx 2$ T/K the field $H_{c2}(0)$ would approach the paramagnetic limit of ≈ ~70 Tesla for MgB$_2$, in which case strong coupling and spin effects should be included in a more general theory based on the Eliashberg equations [17,38].

**Acknowledgements**:

We have greatly benefited from discussions and assistance from P. Manfrinetti, A. Palenzona, M. Rzchowski, A. Godeke, B, Senkowicz, P. Voyles, L. Balicas, A. Lacerda, and G. Boebinger. The work was primary supported through the NSF Nanostructured Materials and Interfaces Materials Research Science and Engineering Center at the University of Wisconsin.

# References

1. Nagamatsu, J., Nakagawa, N., Muranaka, T., Zenitani, Y. & Akimitsu, J. Superconductivity at 39K in Magnesium Diboride. Nature, **410**, 63-64 (2001).
2. Canfield, Paul C. & Crabtree, George W. Magnesium diboride: better late than never. Physics Today, **56**, 34-40 (2003).
3. Dou, S.X., et al. P. Enhancement of the critical current density and flux pinning of $MgB_2$ by nanoparticle SiC doping. Appl. Phys. Lett. **81**, 3419-3421 (2002).
4. Wang, J., et al. High critical current density and improved irreversibility field in bulk $MgB_2$ made by a scalable, nanoparticle addition route. Appl. Phys. Lett. **81**, 2026-2028 (2002).
5. Komori, K, et al. Approach for the fabrication of $MgB_2$ superconducting tape with large in-field transport critical current density Appl. Phys. Letts. **81**, 1047 (2002).
6. Flukiger, R., Suo, H.L., Musolino, N., Beneduce, C., Toulemonde, P. & Lezza, P. Superconducting properties of $MgB_2$ tapes and wires. Physica C **385**, 286-305 (2003).
7. Larbalestier, D.C., et al. Strongly linked current flow in polycrystalline forms of the new superconductor $MgB_2$. Nature **410**, 186-189 (2001).
8. Angst, M., Puzniak, R., Wisniewski, A., Jun, J., Kazakov, S.M., Karpinski, J., Roos, J. & Keller, H. Temperature and field dependence of anisotropy of $MgB_2$. Phys. Rev. Lett. **88**, 1670041-4 (2002).
9. Perkins, G.K, et al., Superconducting Critical fields and anisotropy of a $MgB_2$ single crystal, Supercond. Sci. & Technol. **15**, 1156-1159 (2002).
10. Sologubenko, A.V., Jun, J., Kazakov, S.M., Karpinski, J. & Ott, H.R., Temperature dependence and anisotropy of the upper critical field of $MgB_2$, Phys. Rev. B **65**, 180505-1-4 (2002).
11. Zehetmayer, M., et al. Mixed state properties of superconducting $MgB_2$ single crystals. Phys. Rev. B **66**, 055051-4 (2002).
12. Canfield, P.C., Bud'ko, S.L. & Finnemore, D.K. An overview of the basic physical properties of $MgB_2$. Physica C **385**, 1-7 (2003).
13. Larbalestier, David, Gurevich, Alex, Feldmann, D. Matthew & Polyanskii, Anatolii. High transition temperature superconducting materials for electric power applications. Nature **414**, 368-377 (2001).
14. Schultz, Joel H. The medium temperature superconductor (MTS) design philosophy. IEEE Trans. Appl. Supercond. **13**, 1604-1607 (2003).
15. Grant, P. Rehearsals for prime time. Nature **411**, 532-533 (2001).
16. Liu, Amy Y., Mazin, I.I. & Kortus, Jens. Beyond Eliashberg superconductivity in $MgB_2$: anharmonicity, two-phonon scattering and multiple gaps. Phys. Rev. Lett. **87**, 0870051-4 (2001).
17. Choi, H.J., Roundy, D., Sun, H., Cohen, M.L. & Loule, S.G. The origin of anomalous superconducting properties of $MgB_2$. Nature **418,** 758-760 (2002).
18. Iavarone, M., et al. Two-band superconductivity in $MgB_2$. Phys. Rev. Lett. **89**, 1870021-4 (2002).
19. Schmidt, H., Zasadzinski, J.E., Gray, K.E. & Hinks, D.G. Evidence for two-band superconductivity from break-junction tunneling in $MgB_2$. Phys. Rev. Lett. **88**, 1270021-4 (2002).
20. Chen, X.K., Kostantinovic, M.J., Irwin, J.C., Lawrie, D.D. & Franck, J.P. Evidence for two gap superconductivity in $MgB_2$. Phys. Rev. Lett. **87**, 1570021-4 (2001).
21. Bouquet, F., Wang, Y., Sheikin, I., Plackowski, T., Junod, A., Lee, S. & Tajima, S. Specific heat of single crystal $MgB_2$: A two-band superconductor with different anisotropies. Phys. Rev. Lett. **89**, 257001-4 (2002).
22. Cubitt, R., Levett, S., Bud'ko, S.L., Anderson, N.E. & Canfield, P.C. Experimental evidence for anisotropic double gap behavior in $MgB_2$. Phys. Rev. Lett. **90**, 1570021-4 (2003).
23. Mazin I.I., et al. Superconductivity in $MgB_2$: Clean or dirty? Phys. Rev. Lett. **89**, 1070021-4 (2002).
24. Fietz, W.A. & Webb, W.W. Magnetic properties of some type II superconductors near the upper critical field, Phys. Rev. **161**, 423-433 (1967).
25. Orlando, T.P., McNiff, E.J., Foner, S. & Beasley, M.R. Critical fields, Pauli paramagnetic limiting, and material parameters of $Nb_3Sn$ and $V_3Si$. Phys. Rev. B**19**, 4545-4561 (1979).

26. Ketterson, J.B. & Song, S.N. Superconductivity (Cambridge University Press, Cambridge, 1999).
27. Patnaik, S., et al. Electronic anisotropy, magnetic field-temperature phase diagram and their dependence on resistivity in c-axis oriented $MgB_2$ thin films. Supercond. Sci.& Technol., **14**, 315-319 (2001).
28. Gumbel, A., et al. Improved superconducting properties in nanocrystalline bulk $MgB_2$. Appl. Phys. Lett. **80**, 2725-2728 (2002).
29. Bugoslavsky, Y., et al. Enhancement of high-field critical current density of superconducting $MgB_2$ by proton irradiation. Nature, **411**, 561-563 (2001).
30. Bouquet, F., et al. Unusual effects of anisotropy on the specific heat of ceramic and single crystal $MgB_2$. Physica C **385**, 192-204 (2003).
31. Eom, C.B., et al. Thin film magnesium boride superconductor with very high critical current density and enhanced irreversibility field. Nature, **411**, 558-560 (2001).
32. Bu, S.D., et al. Synthesis and properties of c-axis oriented $MgB_2$ epitaxial films. Appl. Phys. Lett. **81**, 1851-1853 (2002).
33. Braccini, V., et al. Significant enhancement of irreversibility field in clean-limit bulk $MgB_2$, Appl. Phys. Lett. **81**, 4577-4579 (2002).
34. Song, X., et al. Anisotropic grain morphology, crystallographic texture and their implications for flux pinning in $MgB_2$ pellets, filaments and thin films, Supercond. Sci. & Technol. **15**, 511–518 (2002).
35. Ando, Y et al., Resistive upper critical fields and irreversibility lines of optimally doped high-$T_c$ cuprates, Phys. Rev. B **60**, 12475-12479 (1999).
36. Gurevich, A. Enhancement of $H_{c2}$ by nonmagnetic impurities in dirty two-gap superconductors. Phys. Rev. B **67**, 1845151-13 (2003).
37. Koshelev, A.E. & Golubov, A.A. Mixed state in a dirty two-band superconductor: application to $MgB_2$. Phys. Rev. Lett. **90**, 1770021-4 (2003).
38. Golubov, A.A., et al. Specific heat of $MgB_2$ in a one- and a two-band model from first-principle calculations. J. Phys: Condensed Matter. **14**, 1353-1360 (2002).
39. Cava, R. J., Zandbergen, H.W. & Inumaru. K., The substitutional chemistry of $MgB_2$, Physica C **385**, 8-15, (2003).
40. Rowell, John M. The widely variable resistivity of $MgB_2$ samples. Supercond. Sci. & Technol. **16**, R17-R27 (2003).
41. Lindenhovius, J., Hornsveld, E., den Ouden, A., Wessel, W. & ten Kate, H., Powder-in-Tube (PIT) $Nb_3Sn$ conductors for high field magnets, IEEE Trans. on Appl. Supercon. **10**, 975-978 (2000).

**Figure captions:**

**Figure. 1:** Resistive transitions taken in the LANL pulsed field facility on the high resistivity ($\rho(40K)$ = 220 $\mu\Omega$cm) film. Data were taken in several runs on more than one occasion. The horizontal dashed lines correspond to 0.1R($T_c$) and 0.9R($T_c$) defining the irreversibility and the upper critical fields plotted in Figure 2. Figure 1a: data taken at temperatures of 1.43, 2.12, 2.71, 4.06, 4.5, 5.1, 5.95, 8.11, 11.4, 13.95, 17.1, 20.05, 24, 28 and 30.6 K for field applied perpendicular to the film and to the ab plane. Figure 1b: data taken at temperatures of 1.49, 1.99, 2.7, 3.76, 4.41, 7.13, 10.11, 13.33, 15.99, 18.64, 21.8, 24.8, 27.03, 28.85, 29.9 and 30.7 K for field applied parallel to the film and to the ab plane.

Fig. 2: The upper critical fields $H_{c2}(T)$ (black squares) and irreversibility fields $H^*(T)$ (open squares) for high resistivity 220 $\mu\Omega$cm film, and $H_{c2}(T)$ for low resistivity 7 $\mu\Omega$cm film (black triangles). Solid curves show $H_{c2}(T)$ calculated from Equation (2) with the coupling constants, $\lambda_{\sigma\sigma}$ = 0.81, $\lambda_{\pi\pi}$ = 0.28, $\lambda_{\sigma\pi}$ = 0.115 and $\lambda_{\pi\sigma}$ = 0.09 [38]. Figure 2a corresponds to **H**$\perp$ab, for which the theoretical curve is plotted for $D_\pi$ = 0.12$D_\sigma$. Figure 2b corresponds to **H**$\parallel$ab, for which the theoretical curve is plotted for $[D_\pi^{(c)}D_\pi^{(ab)}]^{1/2}$ =0.2$[D_\sigma^{(c)}D_\sigma^{(ab)}]^{1/2}$. The dashed lines show $H_{c2}(T)$ curves for MgB$_2$ single crystals [11].

Fig. 3: Dc resistive transitions at a measuring current density ~1 A/cm$^2$ on bulk samples: the clean one ($\rho$(40 K) = 1 $\mu\Omega$cm, black squares), the sample heated in Mg vapor and then slow cooled ($\rho$(40 K) = 18 $\mu\Omega$cm, black diamonds), and the same sample (aged to $\rho$ = 5 $\mu\Omega$cm, black circles) measured at the NHMFL two months later. In the latter case, the low resistivity and short length of the sample significantly decreased the signal to noise ratio at the highest fields and lowest temperatures, as indicated by the error bars. For lower fields, the error bars are not shown because they were smaller than 1T. In any case, $H_{c2}(T)$ significantly increases as the resistivity increases, the slope $H_{c2}'$ = 1.2 T/K at 25-30 K for the 18 $\mu\Omega$cm sample being almost as high as $H_{c2}'$ for the 220 $\mu\Omega$cm film at **H**$\perp$ab in Figure 2a. The dashed line shows $H_{c2}(T)$ of Nb$_3$Sn.

Fig. 4: Temperature dependence of the anisotropy parameter $H_{c2}^{\parallel}/H_{c2}^{\perp}$ for the 220 $\mu\Omega$cm dirty MgB$_2$ film (solid squares) and 7 $\mu\Omega$cm epitaxial film (open triangles) obtained from the data presented in Figure 2. The solid curve is calculated from Equations (2) and (5) for $D_\pi^{(ab)}$ = 0.09$D_\sigma^{(ab)}$ and $D_\sigma^{(ab)}$ = 3$D_\sigma^{(c)}$.

Fig. 5: Transmission electron micrographs of the two films. a) High resistivity film ($\rho$(40 K) = 220 $\mu\Omega$cm) showing c-axis textured "brick-wall" structure with imperfectly distinguished MgB$_2$ and MgO grains with characteristic size of 10-20 nm. b) Atomic resolution image showing the highly strained and buckled Mg structure, especially along the c-axis. Variable intensity of the Mg bright spots indicates buckling of Mg rows in the direction perpendicular to the plane of the figure. c) More perfect structure of the epitaxial film showing much less evidence for distortion in this low resistivity film ($\rho$(40 K) = 7 $\mu\Omega$cm).

Fig. 6: Comparison of upper critical fields of low-$T_c$ superconductors. Here we show $H_{c2}(T)$ curves for Nb-Ti, high-field Nb$_3$Sn, high resistivity MgB$_2$ film of this work and PbMo$_6$S$_8$ which has the highest H$_{c2}$ for non-cuprate superconductors. The high field performance of MgB$_2$ is superior of that of Nb$_3$Sn both for parallel and perpendicular fields. The marked area shows the temperature and field range where the "medium temperature" superconducting applications may be most effective [14]. The vertical dashed lines show 1 atmosphere boiling points of different cryogenic coolants.

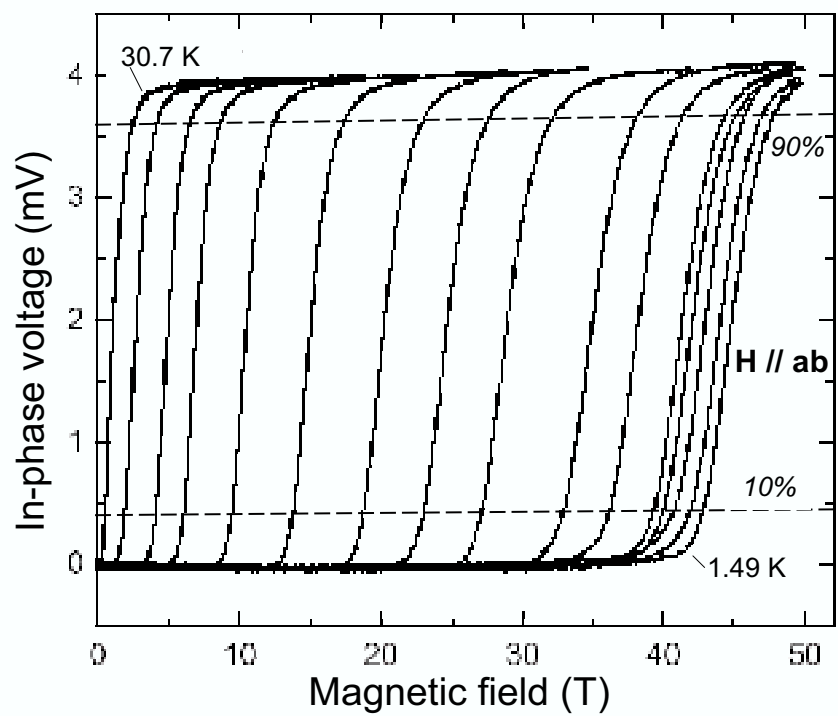

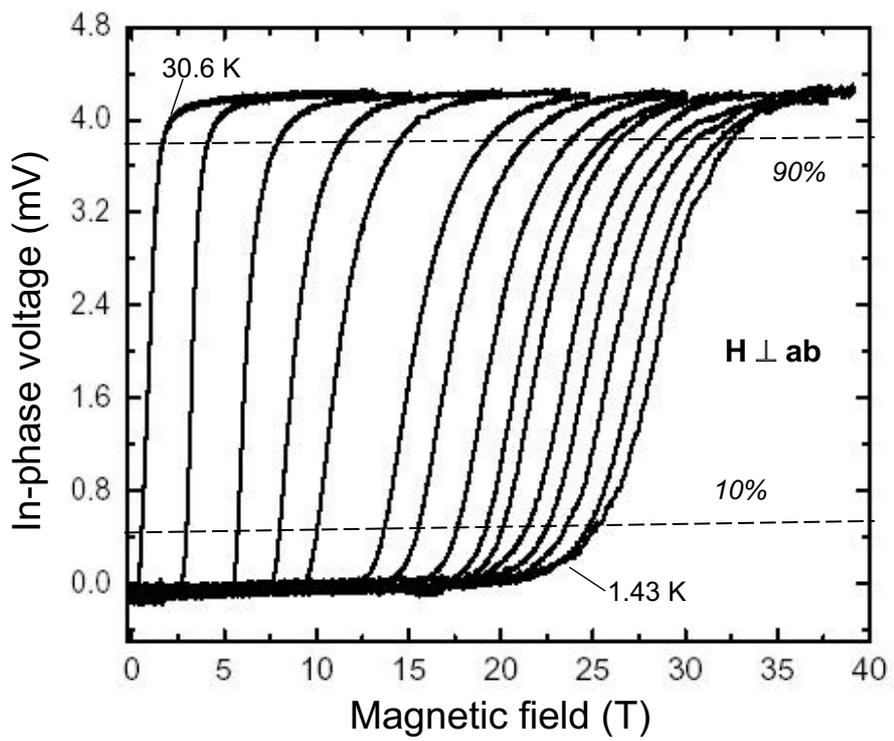

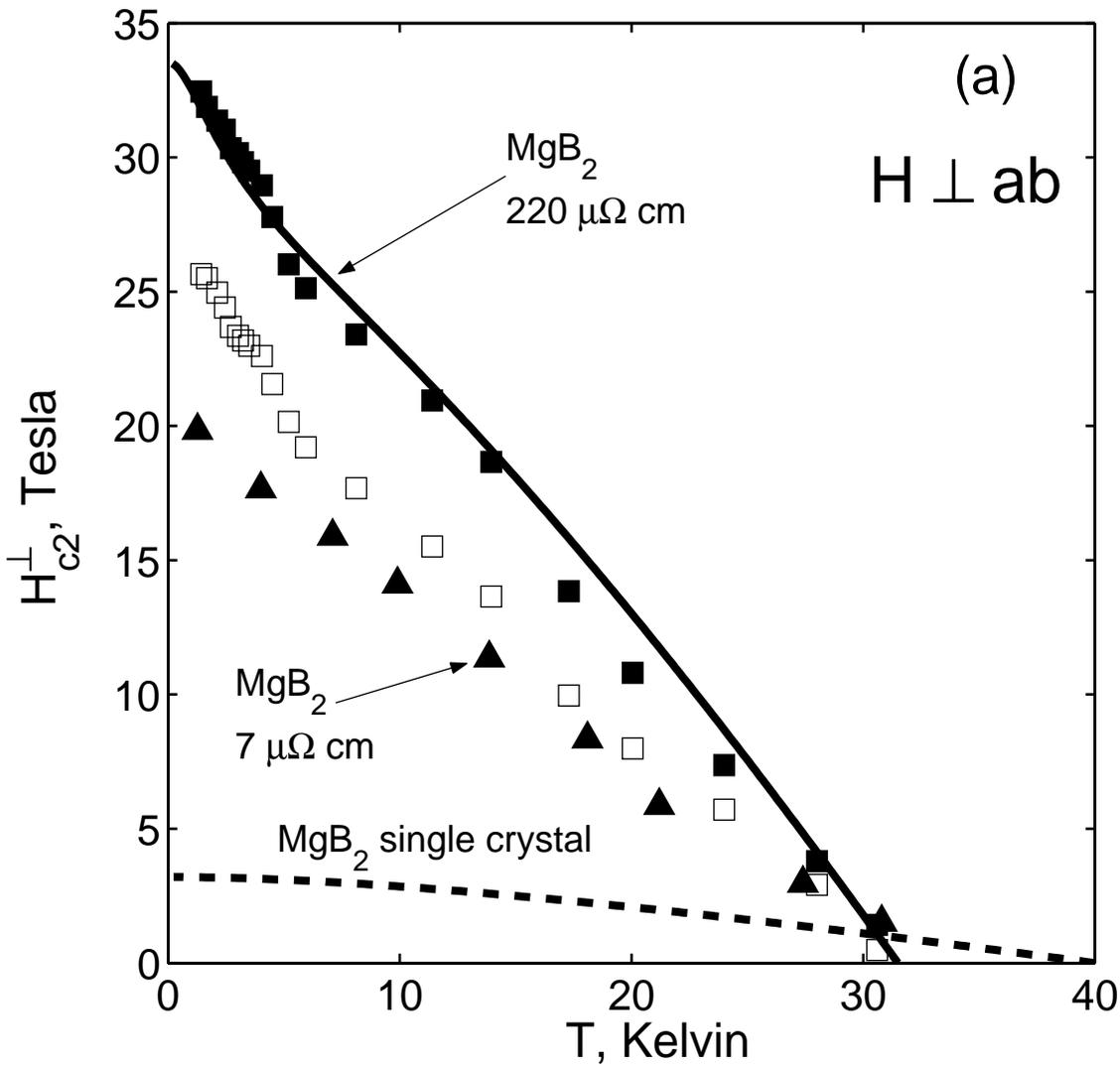

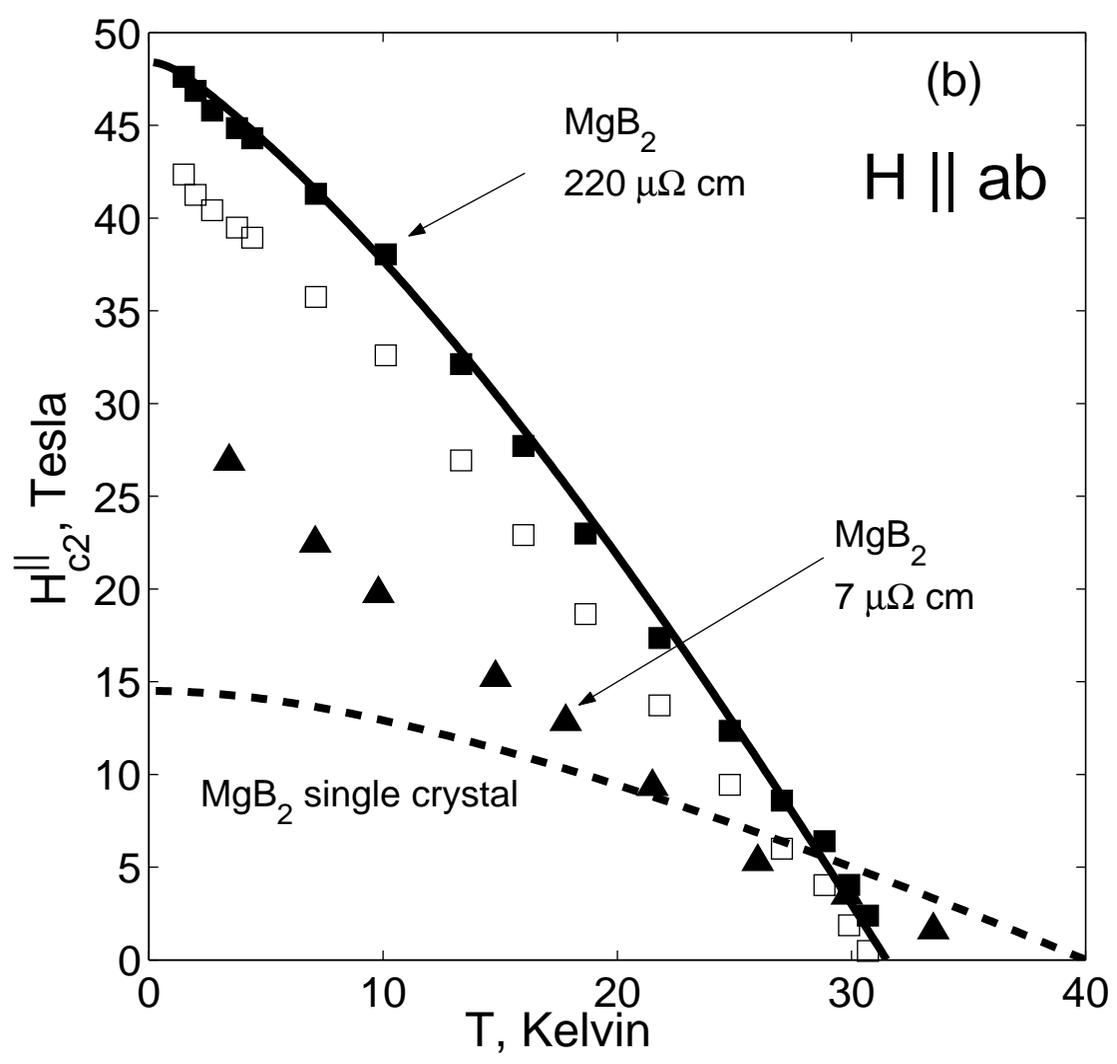

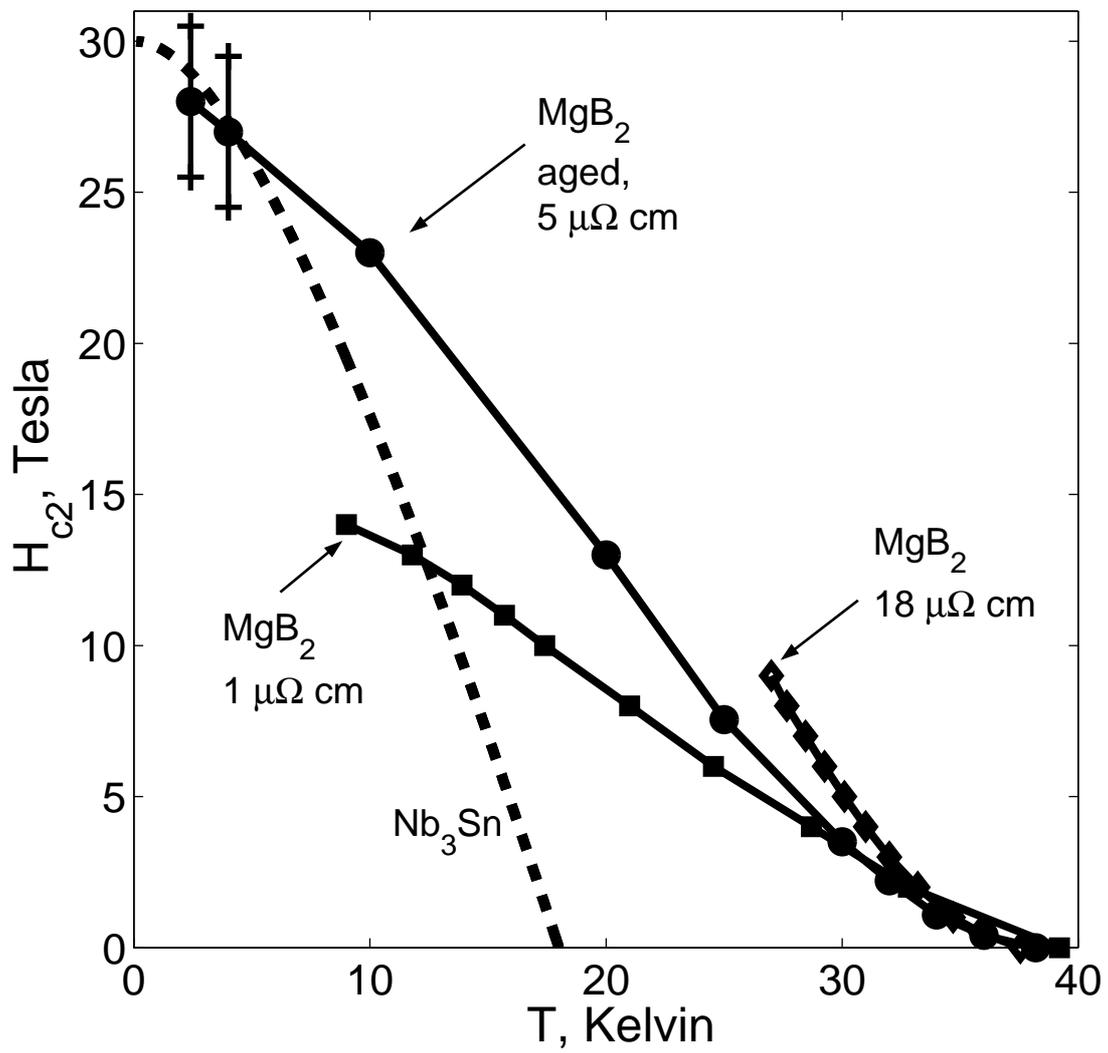

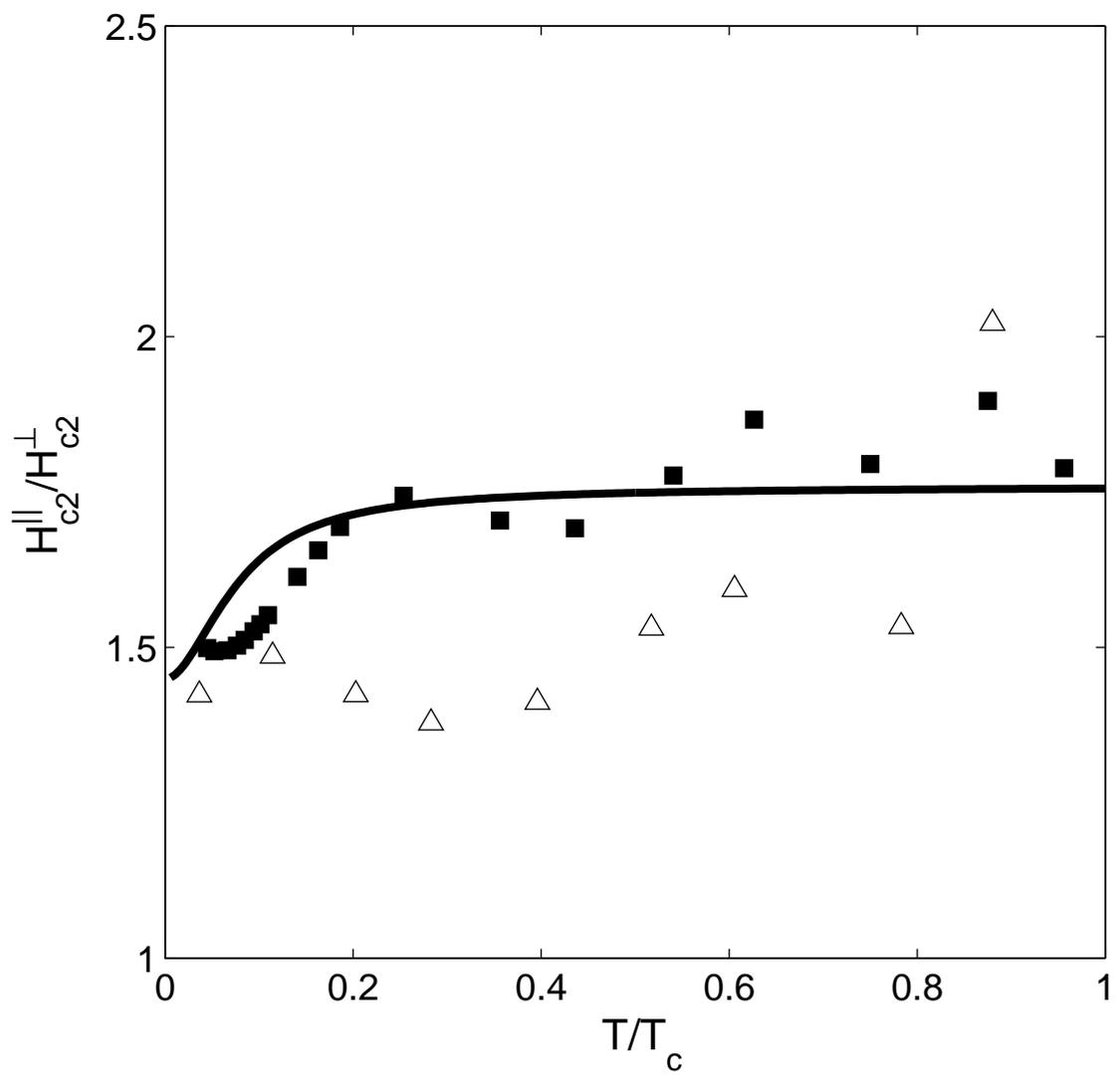

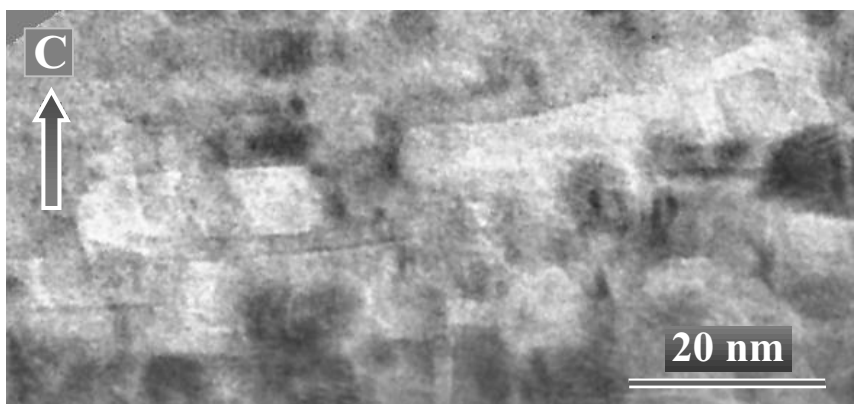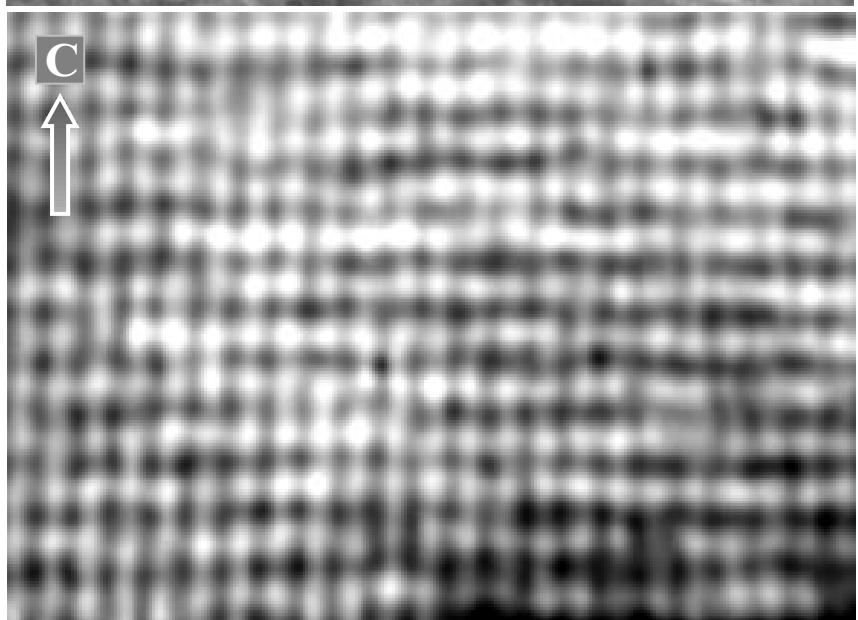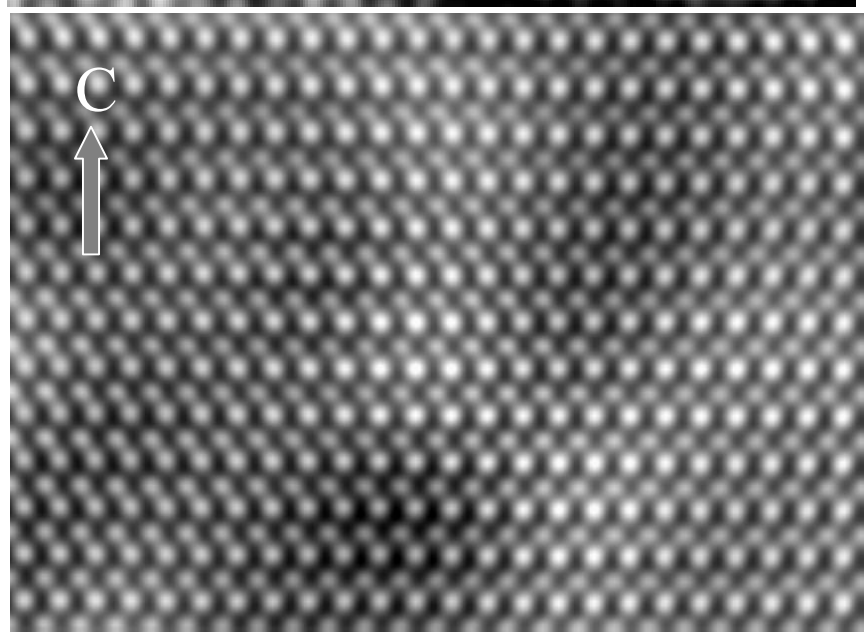

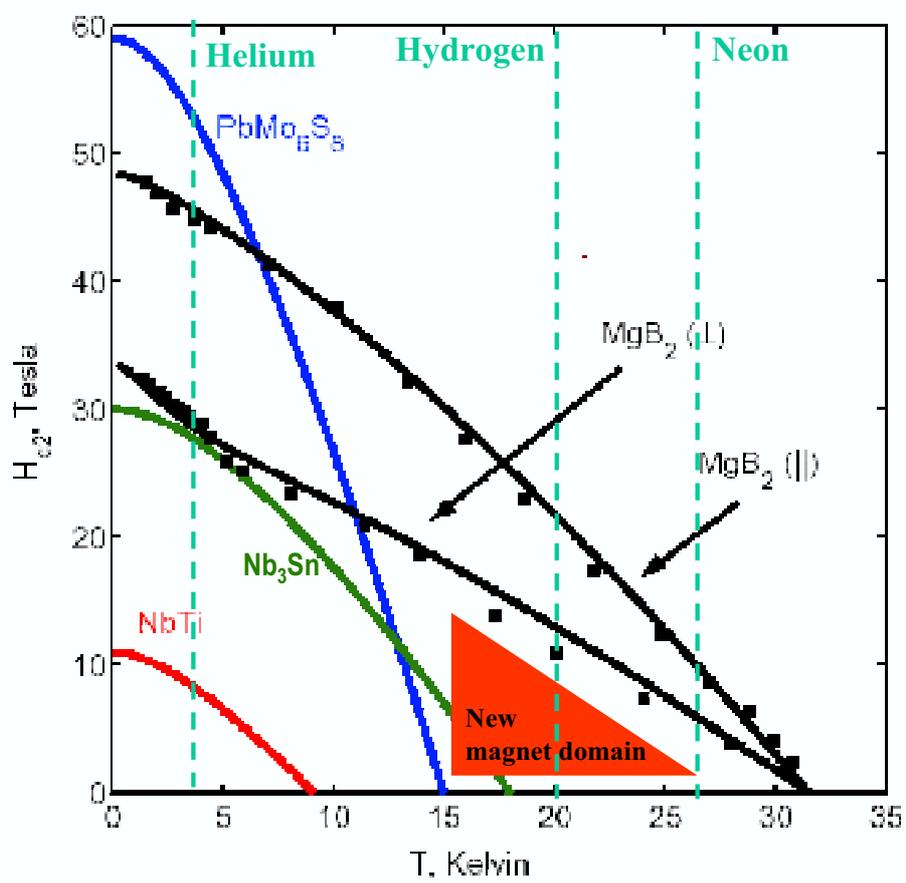